\documentclass[10pt,letterpaper]{article}
\usepackage{mathptmx}
\usepackage[utf8]{inputenc}
\usepackage[letterpaper]{geometry}
\usepackage{fancyhdr}
\usepackage{verbatim}
\usepackage{units}
\usepackage{amsmath}
\usepackage{amssymb}
\usepackage{graphicx}
\usepackage{microtype}
\usepackage[unicode=true,
 bookmarks=false,
 breaklinks=false,pdfborder={0 0 1},backref=section,colorlinks=false]
 {hyperref}

\makeatletter
\@ifundefined{date}{}{\date{}}
\usepackage{changepage}

\usepackage[utf8]{inputenc}  
\usepackage{textcomp,marvosym}

\usepackage{fixltx2e}

\usepackage{amsmath,amssymb}

\usepackage{cite}

\usepackage{nameref,hyperref}

\usepackage[right]{lineno}

\DisableLigatures[f]{encoding = *, family = * }

\usepackage{rotating}

\usepackage[aboveskip=1pt,labelfont=bf,labelsep=period,justification=raggedright,singlelinecheck=off]{caption}

\renewcommand{\@biblabel}[1]{\quad#1.}

\usepackage{lastpage}
\usepackage{fancyhdr}\usepackage{graphicx}\usepackage{epstopdf}
\fancyfootoffset[L]{2.25in}

\renewcommand{\fnum@figure}{Fig ~\thefigure}

\def\and{%
  \end{tabular}%
  \hfill%
  \hspace*{\columnsep}%
  \hfill%
  \begin{tabular}[t]{c}}%
 \def\@maketitle{%
  \newpage%
  \null%
  \vskip 2em%
  \begin{center}%
  \let \footnote \thanks%
    {\LARGE \@title \par}%
    \vskip 1.5em%
    {\large%
      \lineskip .5em%
      \hfill%
      \begin{tabular}[t]{c}%
        \@author%
      \end{tabular}\hfill\null\par}%
    \vskip 1em%
    \vskip 1em%
    {\large \@date}%
  \end{center}%
  \par%
  \vskip 1.5em}%

\@ifundefined{showcaptionsetup}{}{%
 \PassOptionsToPackage{caption=false}{subfig}}
\usepackage{subfig}
\makeatother

\begin{document}
\vspace*{0.35in}
\begin{flushleft}
{\Large \textbf\newline{Swift Two-sample Test on High-dimensional Neural Spiking Data } }

Zhi-Qin John Xu\textsuperscript{1*}, 
Douglas Zhou\textsuperscript{2*},
David Cai\textsuperscript{1,2,3}, 
\\
\bigskip
\bf{1} NYUAD Institute, New York University Abu Dhabi,  Abu Dhabi, United Arab Emirates, 
\\ 
\bf{2}  School of Mathematical Sciences, MOE-LSC and Institute of Natural Sciences, Shanghai Jiao Tong University, Shanghai, P.R. China,  
\\ 
\bf{3} Courant Institute of Mathematical Sciences and Center for Neural Science, New York University, New York, New York, United States of America.
\\ 
\bigskip
* Corresponding Authors: zhiqinxu@nyu.edu, zdz@sjtu.edu.cn
\end{flushleft}

\section*{Abstract}

To understand how neural networks process information, it is important
to investigate how neural network dynamics varies with respect to
different stimuli. One challenging task is to design efficient statistical
approaches to analyze multiple spike train data obtained from a short
recording time. Based on the development of high-dimensional statistical
methods, it is able to deal with data whose dimension is much larger
than the sample size. However, these methods often require statistically
independent samples to start with, while neural data are correlated
over consecutive sampling time bins. We develop an approach to pretreat
neural data to become independent samples over time by transferring
the correlation of dynamics for each neuron in different sampling
time bins into the correlation of dynamics among different dimensions
within the each sampling time bin. We verify the method using simulation
data generated from Integrate-and-fire neuron network models and a
large-scale network model of primary visual cortex within a short
time, \emph{e.g.}, a few seconds. %
{} Our method may offer experimenters to use the advantage of development
of statistical methods to analyze high-dimensional neural data.

\section*{Introduction}

A brain can process information swiftly (in a few hundred milliseconds)
and robustly \cite{chittka2009speed,uchida2006seeing,mainen2006behavioral,abraham2004maintaining,rousselet2003animal,roitman2002response},
however, the understanding of behaviors of the brain within a short
time remains a great theoretical and experimental challenge. There
are many experimental observations suggesting that neural populations
in the brain containing tens to thousands of neurons are correlated
to perform elementary cognitive operations \cite{zandvakili2015coordinated,harris2015long,schneidman2006weak,shlens2006structure}.
The dimension $p$ of the recorded neural data, \emph{i.e.}, the number
of neurons that are simultaneously recorded under a brain state, could
be as large as several hundreds, whereas the sample size $n$ of the
data, \emph{i.e.}, the number of sampling bins in a short recording
time, could be much smaller than $p$ (``large $p$, small $n$'').
The exponentially growing data has created urgent needs to design
methods that can perform multivariate analysis of neural data within
a short recording time \cite{brown2004multiple,buzsaki2004large,chapin2004using,buzsaki2012origin,shimazaki2012state,trousdale2012impact,cunningham2014dimensionality,berenyi2014large,khodagholy2015neurogrid,xu2016dynamical,xu2018dynamical,xu2018maximum}. 

In this work, we aim to design statistical methods to discriminate
two different stimuli from the recordings of activities of hundreds
of neurons in a short time, \emph{e.g.}, hundreds of milliseconds
to a few seconds. To get enough statistical power in a discrimination
analysis, it usually requires neural activities of many identical
trials, such as principle component analysis (PCA). However, the responses
of a neural network vary significantly on nominally identical trials.
This variation is suspected especially to be true for tasks that involve
internal states, say, attention, decision-making and more \cite{cunningham2014dimensionality}.
Therefore, averaging responses across trials may obscure the analysis,
and single-trial analysis with statistical power is therefore essential. 

In our setting, for an experiment trial, we have two population recordings
(a referential one and a test one), such as spike trains, in response
to two stimuli. The goal of our analysis is to discriminate whether
the underlying test stimulus comes from the same source as the referential
one based on the two population recordings in the experimental trial.
We simplify our analysis by testing the null hypothesis that the mean
value of the test sample equals the mean value of the referential
sample. To incorporate information of correlation, statistically,
we can apply high-dimensional two-sample test methods (See Methods.)
that account for correlation structure to detect the difference of
neural activities. Recently, there is great progress in high-dimensional
two-sample test methods to deal with samples of ``large $p$, small
$n$'' \cite{chen2010two,srivastava2008test,lopes2011more,gretton2012kernel,tony2014two,reddi2015high}.
Typically, statistical methods including these high-dimensional two-sample
test methods require statistically independent sampling to start with.
However, neural activity, such as spike train, is dependent in the
following sense. One data point is defined by the neural firings in
one sampling time bin. The decay time of neural dynamics is dozens
of milliseconds, \emph{e.g.}, $\unit[20]{ms}$, then the consecutive
data points are not statistically independent. We refer to ``\emph{dependent
samples}'' as the samples that are correlated over time. The problem
of dependent samples is a large challenge of big data analysis \cite{fan2014challenges},
because except for neural data, there are many types of dependent
samples of ``large $p$, small $n$'', \emph{e.g.}, fMRI and time
course microarray data. However, there are still few studies about
how to deal with dependent samples \cite{han2013principal,fan2014challenges}.
It remains unclear of how to perform a two-sample test for dependent
samples. %

In this work, we propose a method based on averaging and correlation
transferring (ACT for short, see Results for details.) to pretreat
neural dependent samples to become independent ones. The idea of ACT
method is as follows. A time window, which is larger than correlation
time length of samples, is selected to partition samples over time.
The mean value of samples in each time window is calculated to get
an averaged series. Then, each dimension of the averaged series splits
into two dimensions which are consist of data points of odd positions
and even positions in the averaged series, respectively. Through ACT
method, data points in each dimension of new series are independent.
The ACT method enables statistical two-sample test methods to be applied
to deal with neural recordings from hundreds of milliseconds to seconds.
Comparing to the naive method of discarding partial data, the ACT
method compensate the power of high-dimensional statistical methods
by increasing the dimension of data. In our work, the discrimination
analysis is performed by only the CQ method \cite{chen2010two} (See
 Methods, we call this method CQ for short.), while other high-dimensional
two-sample test methods \cite{srivastava2008test,lopes2011more,gretton2012kernel,tony2014two,reddi2015high}
can be applied in the same procedure. 

We perform two-sample tests on samples that are obtained as spike
trains from the conductance-based integrate-and-fire (I\&F) type neural
networks. It has been shown in experiment that I\&F models can statistically
faithfully capture the response of cortical neurons under in-vivo-like
currents in terms of both firing dynamics and subthreshold membrane
dynamics \cite{rauch2003neocortical}. To show the advantage of incorporating
correlations, we compare the high-dimensional two-sample test with
the Student t-test (TT) (See Methods.), in which neurons are assumed
to be independent. Results show that, based on the ACT method, the
high-dimensional two-sample test method is much better to detect difference
in stimuli from high-dimensional neural spike trains than the TT method
in all scanned dynamical regimes exemplified by two scenarios: one
is that we performed the two-sample tests for $\unit[2]{s}$ recording
of both referential and test samples; the other is that by only using
the referential sample to estimate the variance of statistics, we
performed the two-sample tests for the case that the recording time
for a test data is very short, \emph{i.e.}, $\unit[360]{ms}$, while
the recording time for a referential data is $\unit[4]{s}$. We have
also used the high-dimensional two-sample test method to build one
tuning curve of population neurons from a large-scale network model
of primary visual cortex. This tuning curve is much sharper than a
tuning curve of firing rate and a tuning curve that uses the firing
rates of two neurons \cite{samonds2003cooperation}. The sharper tuning
curve indicates that the high-dimensional two-sample test is potentially
more sensitive with respect to different inputs. Therefore, our results
show that by incorporating the information of correlations among neurons,
the high-dimensional two-sample test based on ACT method can detect
difference in stimuli from the neural activities recorded in very
short time.

The remainder of this paper is organized as follows. First, we show
how to understand the results for a two-sample test of multiple trials.
Second, we describe ACT method in details. Third, we show ACT method
is applicable for data of $\unit[2]{s}$ recording of I\&F neurons
in most dynamical regimes. Fourth, we show ACT method is applicable
for the case that the recording time of I\&F neurons for a test data
is only $\unit[360]{ms}$ while the recording time for a referential
data is $\unit[4]{s}$. Fifth, we apply the high-dimensional two-sample
test method to build one tuning curve of population neurons from a
large-scale network model of primary visual cortex. Finally, we present
our discussion and conclusion.

\section*{Results}

The systems we study here are conductance-based I\&F type neural networks
{[}See Eq. (\ref{eq: IF 02}) in Methods.{]}. We record binary spike
trains with sampling time bin size of $\unit[0.5]{ms}$ from trajectories
of neurons obtained by evolving system Eq. (\ref{eq: IF 02}) numerically.
If a neuron fires during the recorded bin, its recorded value is $1$,
otherwise $0$.

The null hypothesis $H_{0}$ is that the mean values of the referential
and the test samples are the same in every dimension. The significance
level for a two-sample test is selected as $5\%$. For the TT method,
we perform a two-sample test for each dimension and accept the null
hypothesis $H_{0}$ when all dimensions pass the test. To keep the
confidence level $95\%$, we set the significance level for each dimension
in the TT method as $0.05/p$ (See Methods.), where $p$ is the dimension
of data. 

\subsection*{Rejection fraction }

\begin{figure}
\begin{centering}
\includegraphics[scale=0.45]{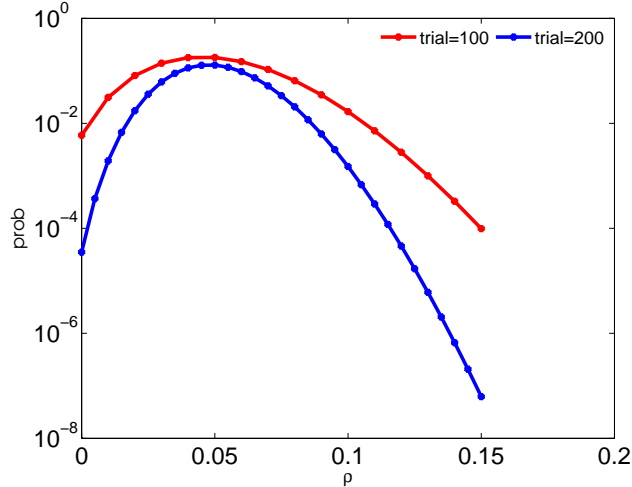}
\par\end{centering}
\caption{{\bf Probability of rejection fractions.} The plot is the probability
of $N\rho$ events rejected against the rejection fraction $\rho$
under the null hypothesis $H_{0}$ {[}Eq. (\ref{eq:Profp}){]} and
the significance level $0.05$. The red and blue lines are the case
of $100$ trials and $200$ trials, respectively.\label{fig:rejprob}}
\end{figure}

Under the null hypothesis $H_{0}$, we can theoretically obtain the
distribution of the statistic of CQ two-sample test method (see Methods).
By selecting the significance level of $5\%$, we can define a confidence
interval (CI) to which the value of the CQ statistic would belong
with probability of $95\%$ under the null hypothesis $H_{0}$. During
a test of one trial, we would reject $H_{0}$ with significance level
of $5\%$ if the value of the statistic is beyond the CI. If there
are multiple trials, what we get is a rejection fraction of $H_{0}$.
The understanding of this rejection fraction is as follows. Since
we have selected the significance level, we can calculate the probability
of every rejection fraction $\rho$ in $N$ trials under $H_{0}$:
\begin{equation}
Prob(f_{rej}=\rho|H_{0},N)=C_{N}^{N\rho}0.05{}^{N\rho}(1-0.05)^{N(1-\rho)},\label{eq:Profp}
\end{equation}
where $C_{N}^{N\rho}$ comes from the selection of $N\rho$ terms
from all the possible $N$ choices. Fig. \ref{fig:rejprob} displays
the corresponding probability of a rejection fraction $\rho$ under
$H_{0}$, \emph{i.e.}, $Prob(f_{rej}=\rho|H_{0},N)$. In the case
of $100$ trials (red), for example, $Prob(\rho=0.15|H_{0},N=100)$
is more than three orders of magnitude smaller than $Prob(\rho=0.05|H_{0},N=100)$.
Hence, it is very significant to reject $H_{0}$ for the rejection
fraction of $0.15$. The function curve of $Prob(f_{rej}=\rho|H_{0},N)$
is sharper in the case of $200$ trials (blue), \emph{e.g.}, $Prob(\rho=0.15|H_{0},N=200)$
is more than six orders of magnitude smaller than $Prob(\rho=0.05|H_{0},N=200)$.
Therefore, it is more significant to reject $H_{0}$ by a rejection
fraction of $0.15$ for the case of $200$ trials than that of $100$
trials. The rejection fraction in an experiment of multiple trials
indicates the significance of rejecting $H_{0}$. Fig. \ref{fig:rejprob}
also shows that a higher rejection fraction (larger than $0.05$)
makes the rejection of $H_{0}$ more significant. Therefore, when
stimuli underlying the test and the referential samples are different,
a higher rejection fraction is better to detect the difference of
these two stimuli. Note that when the test stimulus is the same as
the referential one, the rejection fraction of $0.05$ is the most
probable outcome, which is a consequence of the selected significance
level of $0.05$. Since the rejection fraction can reflect the discriminability
of a two-sample test method, in the following examples, we focus on
the rejection fraction. 

\subsection*{ACT pretreatment for neural data}

In this section, we will elaborate the ACT pretreatment for neural
data. As mentioned above, because of neural dynamics, the consecutive
recording data points of neural activity are not statistically independent.
However, high-dimensional two-sample test methods typically require
statistically independent sampling to start with \cite{srivastava2008test,lopes2011more,gretton2012kernel,tony2014two,reddi2015high,chen2010two}.
For example of CQ method, the CQ statistic $Q_{n}\equiv T_{n}/\hat{\sigma}_{n}$,
where $\hat{\sigma}_{n}$ is defined in Eq. (\ref{eq:sig-1}) in Methods
and
\begin{eqnarray}
T_{n} & = & \frac{\sum_{i\neq j}^{n_{1}}\mathbf{X}_{1i}^{T}\mathbf{X}_{1j}}{n_{1}(n_{1}-1)}+\frac{\sum_{i\neq j}^{n_{2}}\mathbf{X}_{2i}^{T}\mathbf{X}_{2j}}{n_{2}(n_{2}-1)}-\frac{2\sum_{i=1}^{n_{1}}\sum_{j=1}^{n_{2}}\mathbf{X}_{1i}^{T}\mathbf{X}_{2j}}{n_{1}n_{2}},\label{eq:Tnpart-1}
\end{eqnarray}
 $\mathbf{X}_{1}\in\mathbf{R}^{p\times n_{1}}$ and $\mathbf{X}_{2}\in\mathbf{R}^{p\times n_{2}}$
are the referential and the test samples, respectively, $p$ is the
dimension size of samples, $n_{1}$ and $n_{2}$ are the sample sizes
of the referential sample $\mathbf{X}_{1}$ and the test sample $\mathbf{X}_{2}$,
respectively. Under the null hypothesis $H_{0}$, when the data points
of $\mathbf{X}_{1}$ and $\mathbf{X}_{2}$ are sampled statistically
independently, as $p$ tends to infinity, the statistic $Q_{n}$ would
tend to the standard Gaussian distribution, by which the two-sample
test can be performed. Under the null hypothesis $H_{0}$, it can
be shown that the expected value $E(T_{n})=||\mu_{1}-\mu_{2}||_{2}^{2}$
is $0$, where $\mu_{1}$ and $\mu_{2}$ are means of the referential
and the test samples, respectively. We consider a case that samples
are dependent over time, such as $E(X_{k,j}X_{k,j+1})\neq E(X_{k,j})E(X_{k,j+1})$,
where $k=1,2$, $X_{k,j}$ and $X_{k,j+1}$ are a pair of two consecutive
sampling data points of $\mathbf{X}_{k}$, $\mbox{\ensuremath{E(x)}}$
denotes the expectation of stochastic variable $x$. Since $X_{k,j}X_{k,j+1}$
is a part of $T_{n}$ and $E(X_{k,j}X_{k,j+1})$ under dependent sampling
deviates from $E(X_{k,j})E(X_{k,j+1})$ of independent sampling, as
a result, $E(T_{n})$ would also have a deviation from $0$ even the
null hypothesis $H_{0}$ holds. Since the expected value $E(T_{n})$
of the null hypothesis $H_{0}$ is $0$, the deviation of $E(T_{n})$
caused by dependent sampling cannot be dominated. Then, the distribution
of $Q_{n}$ would deviate from the standard Gaussian distribution
under the null hypothesis $H_{0}$. Hence, because of the bias induced
by the correlation in samples over time, the CQ method is not applicable
for dependent samples.

We observe that the correlation over dimensions would not affect $T_{n}$.
To apply two-sample test methods in neural data, our idea is to transfer
the correlation between data points over time into correlation over
dimensions. For the case of one neuron, its spike train is recorded
as a time series $(x_{1},x_{2},\cdots,x_{n})\in\{0,1\}^{n}$, where
$n$ is the total recorded bins. If the correlation length of the
time series is only one, \emph{i.e.}, $E(x_{i}x_{i+1})\neq E(x_{i})E(x_{i+1})$
and $E(x_{i}x_{i+j})=E(x_{i})E(x_{i+j})$ for $i\leq n-j$ and $j\geq2$.
We can turn the time series into a new two-dimensional series (Here,
we assume $n$ is an even number. If $n$ is an odd number, we discard
the last data point.)
\begin{equation}
Y=\left[\begin{array}{ccccc}
x_{1} & x_{3} & \cdots & x_{n-3} & x_{n-1}\\
x_{2} & x_{4} & \cdots & x_{n-2} & x_{n}
\end{array}\right],\label{eq:newSY}
\end{equation}
in which every dimension of $Y$ is an independent series over time.
In general, we don't know the correlation length of a neural time
series. Since the correlation time length of an aperiodic neural dynamical
is usually around dozens of milliseconds, %
{} we can select a time window $\Delta T$ which is much larger than
the neural correlation time length to partition the recording time.
Multiple data points exist in every $\Delta T$ window. For example,
if the sampling time bin size is $\unit[0.5]{ms}$ and $\Delta T=\unit[60]{ms}$,
then, each $\Delta T$ window has $120$ data points. The mean value
of these data points is calculated to represent each $\Delta T$ window.
Since the correlation length of a neural time series is smaller than
the selected $\Delta T$ , if two data points have time distance larger
than $\Delta T$, they are effectively statistically independent.
Hence, the correlation length of the new time series is only one.
We can perform the same process as Eq. (\ref{eq:newSY}) to split
every dimension into a two-dimensional series. If the original series
has dimension $p$, the dimension of the new series is $2p$ as follows
\begin{equation}
\left[\begin{array}{ccccc}
x_{11} & x_{13} & \cdots & x_{1,n-3} & x_{1,n-1}\\
x_{12} & x_{14} & \cdots & x_{1,n-2} & x_{1,n}\\
x_{21} & x_{23} & \cdots & x_{2,n-3} & x_{2,n-1}\\
x_{22} & x_{24} & \cdots & x_{2,n-2} & x_{2,n}\\
\vdots & \cdots &  & \vdots & \vdots
\end{array}\right].\label{eq:SY2}
\end{equation}
 We refer this process as the averaging and correlation transferring
(ACT) method. Next, we would address the following problem of ACT:

(i) There are still some extra correlations, such as $x_{2}$ and
$x_{3}$ in Eq. (\ref{eq:newSY}), which we refer to as antidiagonal
correlation. In the estimator of the variance part of CQ statistic
$Q_{n}$ {[}Eq. (\ref{eq:sig-1}) in Methods{]}, the antidiagonal
correlation is not allowed, either. What is the influence of the antidiagonal
correlation? 

(ii) There are infinite values larger than the neural correlation
length. How can we select a proper time window $\Delta T$?

\begin{figure}
\begin{centering}
\subfloat[Same]{\begin{centering}
\includegraphics[scale=0.28]{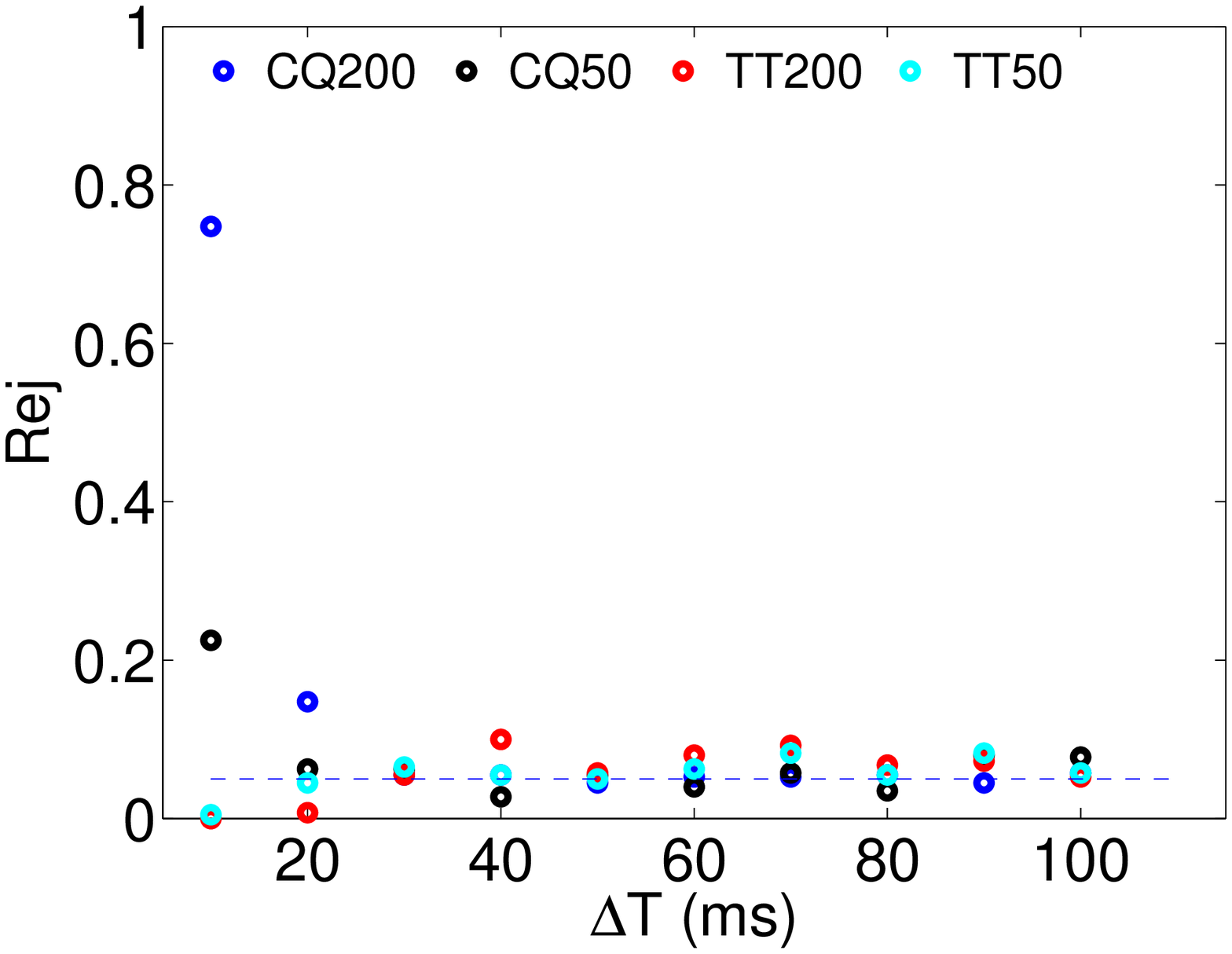}
\par\end{centering}
}\subfloat[Different]{\begin{centering}
\includegraphics[scale=0.28]{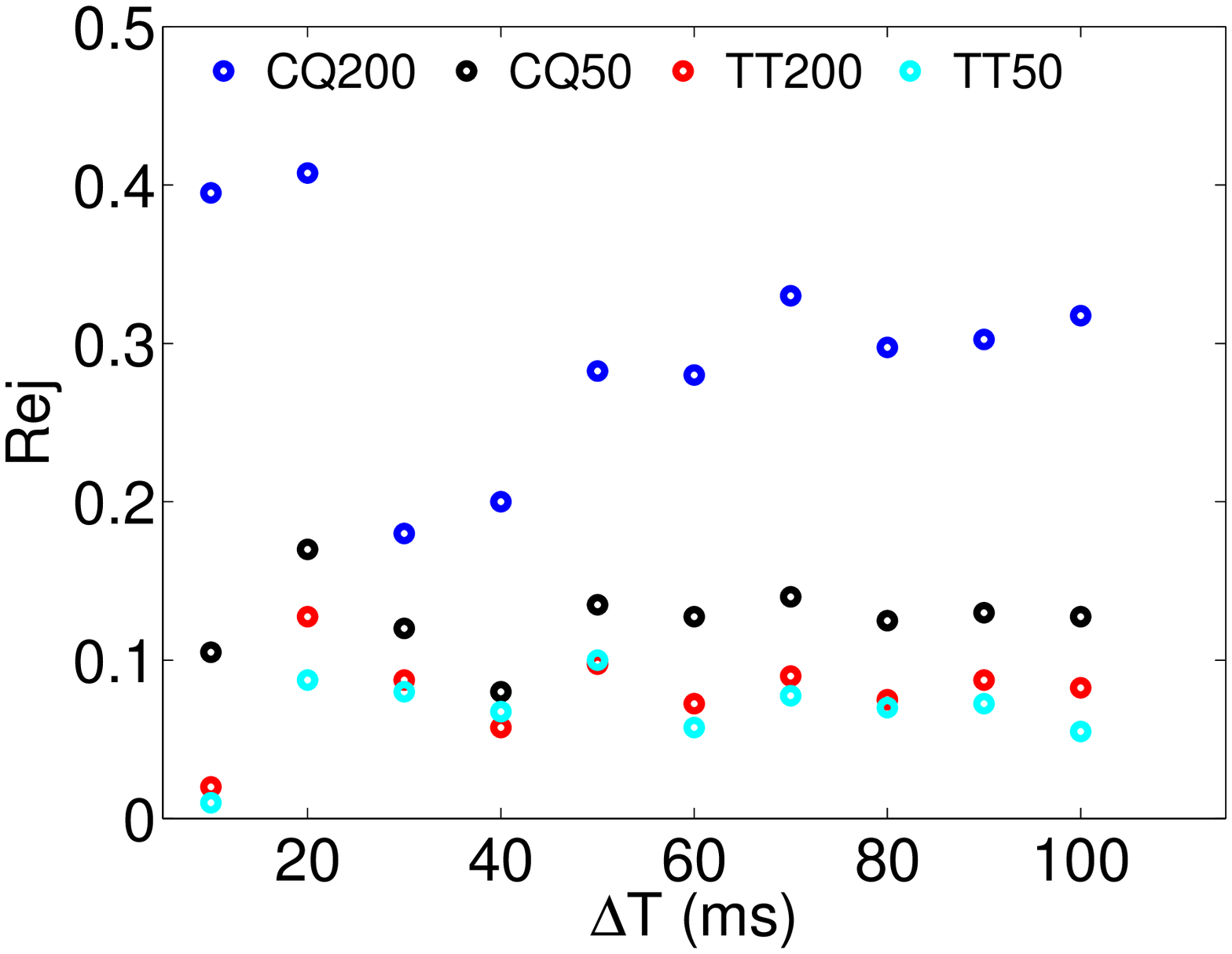}
\par\end{centering}
}
\par\end{centering}
\begin{centering}
\subfloat[$\Delta T=10\unit{ms}$ ]{\begin{centering}
\includegraphics[scale=0.28]{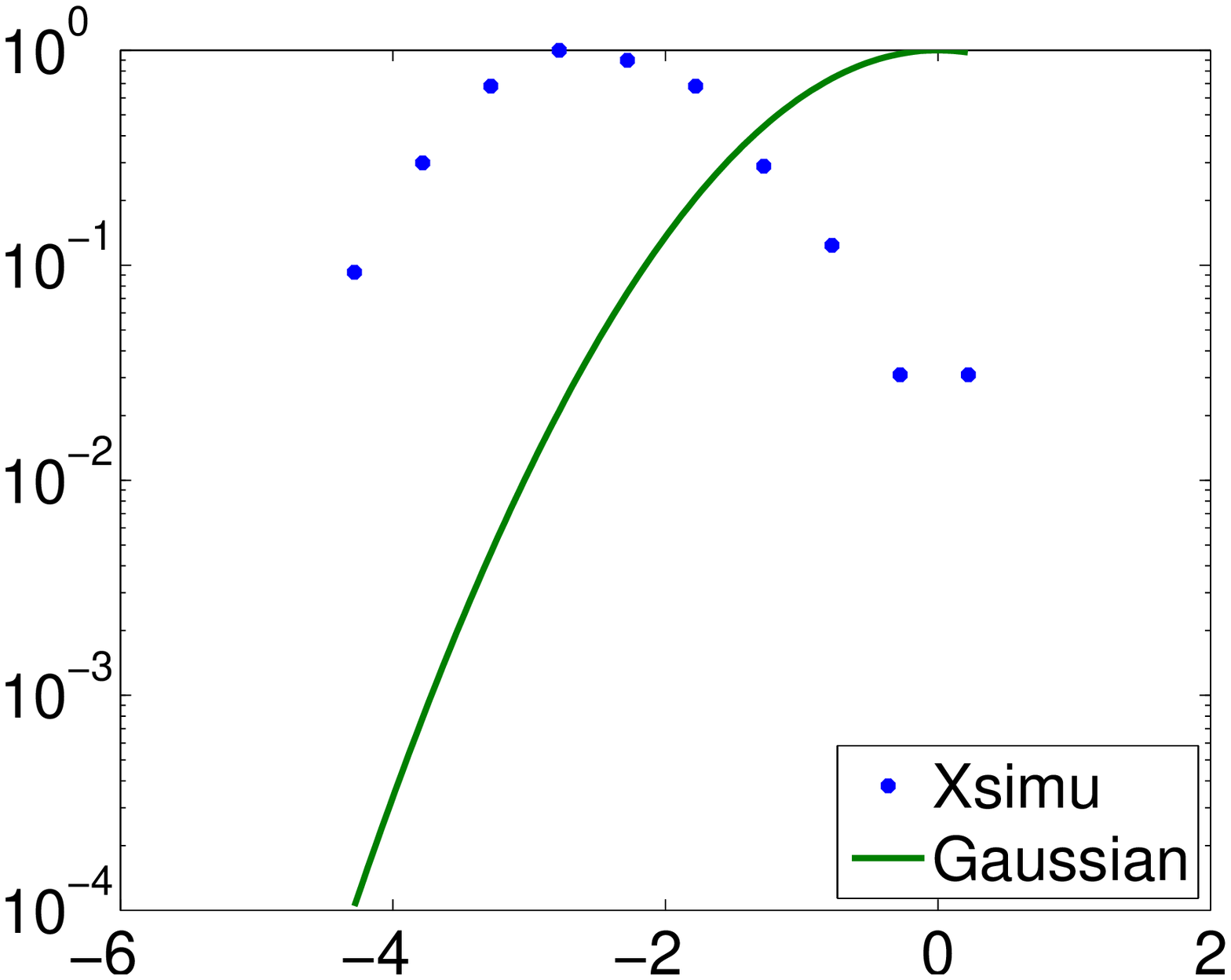}
\par\end{centering}
}\subfloat[$\Delta T=60\unit{ms}$]{\begin{centering}
\includegraphics[scale=0.28]{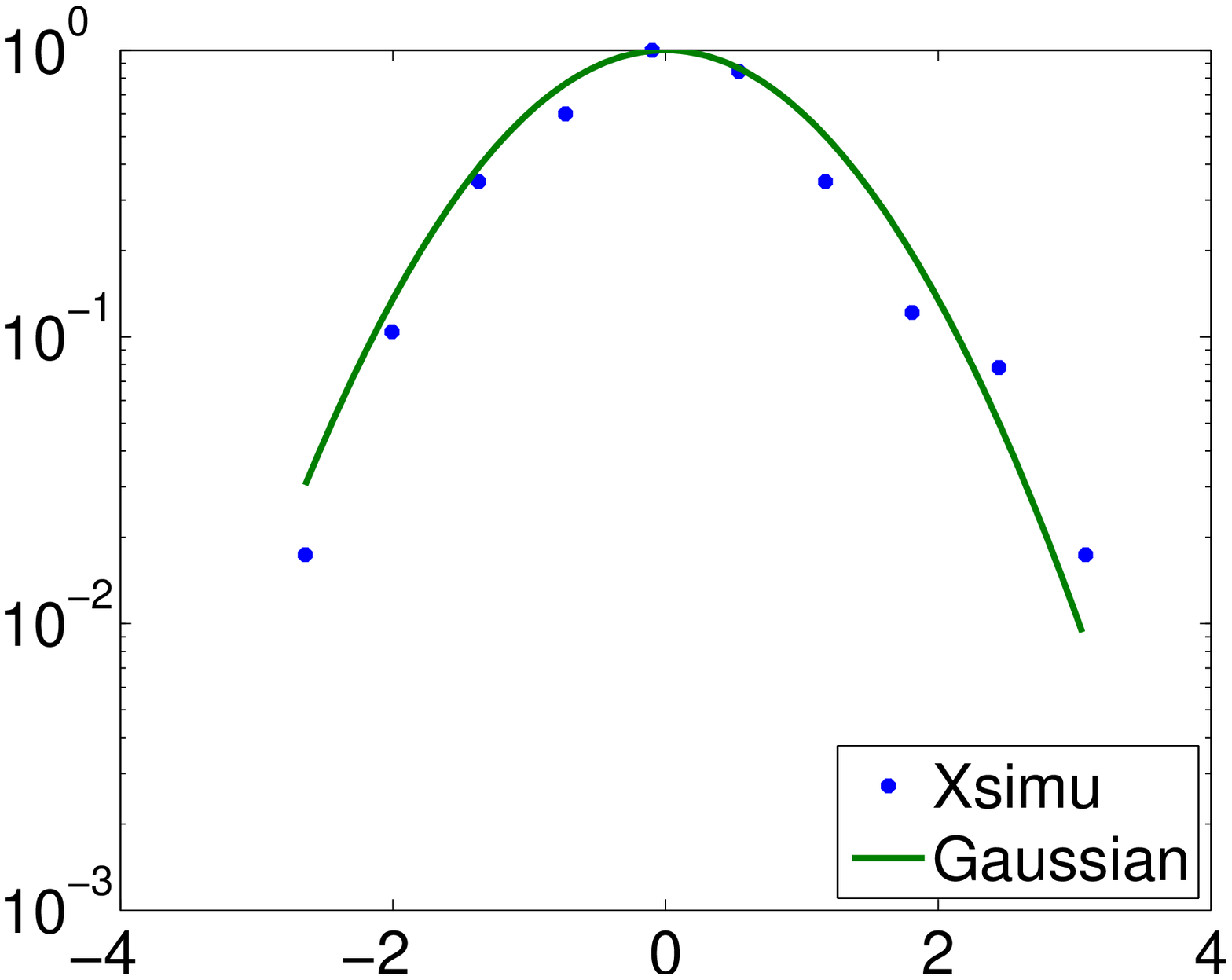}
\par\end{centering}
}
\par\end{centering}
\caption{{\bf Test results for different time window $\Delta T$.} The network
consists of $160$ excitatory and $40$ inhibitory neurons with random
connections with probability $0.1$ driven by a Poisson input. The
coupling strength is $s=0.001\unit{ms^{-1}}$ (the corresponding excitatory/inhibitory
postsynaptic potential is$\sim0.1\unit{mV}$). The input parameters
are chosen $\mu=\unit[0.7]{\unit{ms^{-1}}}$ (Poisson input rate)
and $f=\unit[0.008]{\unit{ms^{-1}}}$ (Poisson input magnitude). The
sampling time bin size is $\unit[0.5]{ms}$. Every trial requires
$\unit[2]{s}$ recording for each stimulus. The mean firing rate is
around $\unit[20]{Hz}$. We apply  two-sample test methods for $400$
trials in four situations, CQ with $200$ neurons (blue), CQ with
$50$ neurons (black), TT with $200$ neurons (red) and TT with $50$
neurons (cyan). For (a) and (b), the plot is the rejection fraction
against the time window $\Delta T$. (a) The test stimuli are the
same as the referential one. The dashed line is $y=0.05$. (b) The
test stimuli are $1.8\%$ larger in Poisson input magnitude than the
referential stimuli. (c) and (d) are the distributions of the statistic
$Q_{n}$ for the case of $200$ neurons  in (a) for $\Delta T=\unit[10]{ms}$
and $\Delta T=\unit[60]{ms}$, respectively. Blue dots are the numerical
results and green lines are the standard Gaussian distributions, where
both are normalized by the corresponding maximum values. \label{fig:MDEPara}}
\end{figure}

We first use a numerical example to study the influence of different
$\Delta T$. Fig.  \ref{fig:MDEPara} displays a numerical example
of scanning $\Delta T$ while applying ACT method to analyze data
recorded from an I\&F neural network. The network consists of $160$
excitatory and $40$ inhibitory neurons with random connections of
probability $0.1$ driven by a Poisson input. Every trial requires
$\unit[2]{s}$ recording time for each stimulus. The mean firing rate
of each neuron is around $\unit[20]{Hz}$. We apply  two-sample test
methods for $400$ trials in four situations, namely, CQ with $200$
neurons (blue), CQ with $50$ neurons (black),  TT with $200$ neurons
(red) and TT with $50$ neurons (cyan). Note that the $50$ neurons
are selected at random.

In Fig.  \ref{fig:MDEPara}a, the test stimulus is the same as the
referential one. When $\Delta T$ is not too small (larger than $\unit[20]{ms}$),
the rejection fractions of all cases are around the selected significance
level $0.05$, which is consistent with the most probable outcome
in Fig.  \ref{fig:rejprob}. We can also examine whether the distribution
of the statistic $Q_{n}$ tends to the standard Gaussian distribution.
When $\Delta T$ is only $\unit[10]{ms}$, the distribution of $Q_{n}$
for the sample of $200$ neurons is far from the Gaussian distribution
(Fig. \ref{fig:MDEPara}c). When $\Delta T$ is larger, such as $\unit[60]{ms}$,
the distribution of $Q_{n}$ for the sample of $200$ neurons fits
the Gaussian distribution well (Fig. \ref{fig:MDEPara}d). These phenomena
indicate that the antidiagonal correlations barely have influence
for a large $\Delta T$. There are two sources of antidiagonal correlation,
one is the consecutive recordings of a neuron, such as $x_{12}$ and
$x_{13}$ in Eq. (\ref{eq:SY2}), the other is the consecutive recordings
of different neurons, such as $x_{22}$ and $x_{13}$ in Eq. (\ref{eq:SY2}).
Here is the reason why the antidiagonal correlations barely have influence:
(i) We select a $\Delta T$ larger than the correlation length of
neuron data, then, the antidiagonal correlations from the consecutive
recordings would be weak comparing to the variance; (ii) The correlation
coefficient between neurons are normally rather weak ($\leq0.1$)
\cite{Cohen2011Measuring}, then the antidiagonal correlations from
the consecutive recordings of different neurons would also be rather
weak comparing to the variance; (iii) In addition, for a data of dimension
$p$, there are many non-zeros in the covariance of Eq. (\ref{eq:SY2}),
\emph{i.e.}, $p$ variances of all dimensions and $p(p-1)$ pairwise
correlations between dimensions. Therefore, the bias induced by antidiagonal
correlations can be dominated. The correlations of neural data over
time do not vanish, whereas ACT method split them into two parts,
one is correlations between dimensions which can be dealt with by
statistical methods, the other is antidiagonal correlations which
can be dominated.

We also found that, when the test stimulus is different from the referential
one, the rejection fraction of $H_{0}$ is not very sensitive to the
parameter $\Delta T$ when $\Delta T$ is sufficiently large. The
expectation of the statistic $Q_{n}$ {[}Eq. (\ref{eq:Qn}) in Methods{]}
in CQ method is crucial to the rejection fraction. Hence, we study
the expectation $E(Q_{n})$ for different $\Delta T$. We can compute
the expectation $E(Q_{n})$ by Eq. (\ref{eq:EQn}) in Methods. We
assume that $\Delta T$ is larger than the correlation time length
of neural data and the total recording time is enough long. We denote
$n$ as the sample size of both two samples with a time window $\Delta T$.
When we use $\Delta T^{\prime}=\alpha\Delta T$, the new sample size
is 
\begin{equation}
n^{\prime}\approx n/\alpha.\label{eq:newN}
\end{equation}
Since the correlation between neurons is small comparing to the variance
\cite{Cohen2011Measuring} and we have assumed that $\Delta T$ is
larger than the correlation length of neuron data, we can assume the
antidiagonal correlation is weak. Under this assumption, the covariance
$\tilde{\Sigma}^{\prime}$ for $\Delta T^{\prime}$ and $\tilde{\Sigma}$
for $\Delta T$ (see  Eq. (\ref{eq:sig-1}) in Methods) have the following
relation (See Appendix for a proof.) 
\begin{equation}
\tilde{\Sigma}^{\prime}\approx\tilde{\Sigma}/\alpha,\label{eq:newS}
\end{equation}
 then substituting Eqs. (\ref{eq:newN}) and (\ref{eq:newS}) into
Eq. (\ref{eq:EQn}), we can obtain the expectation of $Q_{n}^{\prime}$
for $\Delta T^{\prime}$, 
\[
E(Q_{n}^{\prime})\approx2C\frac{(n/\alpha)||\mu_{1}-\mu_{2}||^{2}}{\sqrt{tr(\tilde{\Sigma}^{2}/\alpha^{2})}}=2C\frac{n||\mu_{1}-\mu_{2}||^{2}}{\sqrt{tr(\tilde{\Sigma}^{2})}}=E(Q_{n}),
\]
 where the coefficient $C=1/(4\sqrt{2})$ as defined in Eq. (\ref{eq:EQC})
in Methods. Hence, $\Delta T$ would not affect the test result too
much. But our recording time is finite, $\Delta T$ should not be
selected as a too large value. Numerical simulations also show that
the test result is not very sensitive to the parameter $\Delta T$.
As is shown in Fig. \ref{fig:MDEPara}b, where the test stimuli are
$1.8\%$ larger in Poisson input magnitude than that of the referential
stimuli, the rejection fractions of CQ and TT method do not change
a lot when $\Delta T$ is not too small (larger than $\unit[40]{ms}$).
We can also see that CQ have much higher rejection fractions than
TT, and the results of CQ are better of $200$ neurons (blue dots)
than those of $50$ neurons (black dots). 

Since we have shown that the test result is not very sensitive to
the parameter $\Delta T$, in the following, we will fix $\Delta T$
as $\unit[60]{ms}$ and show that this data pretreatment is applicable
in most neural dynamical regimes.

\subsection*{Numerical experiments on various neural dynamical regimes}

\begin{figure*}
\begin{centering}
\subfloat[Same]{\begin{centering}
\includegraphics[scale=0.21]{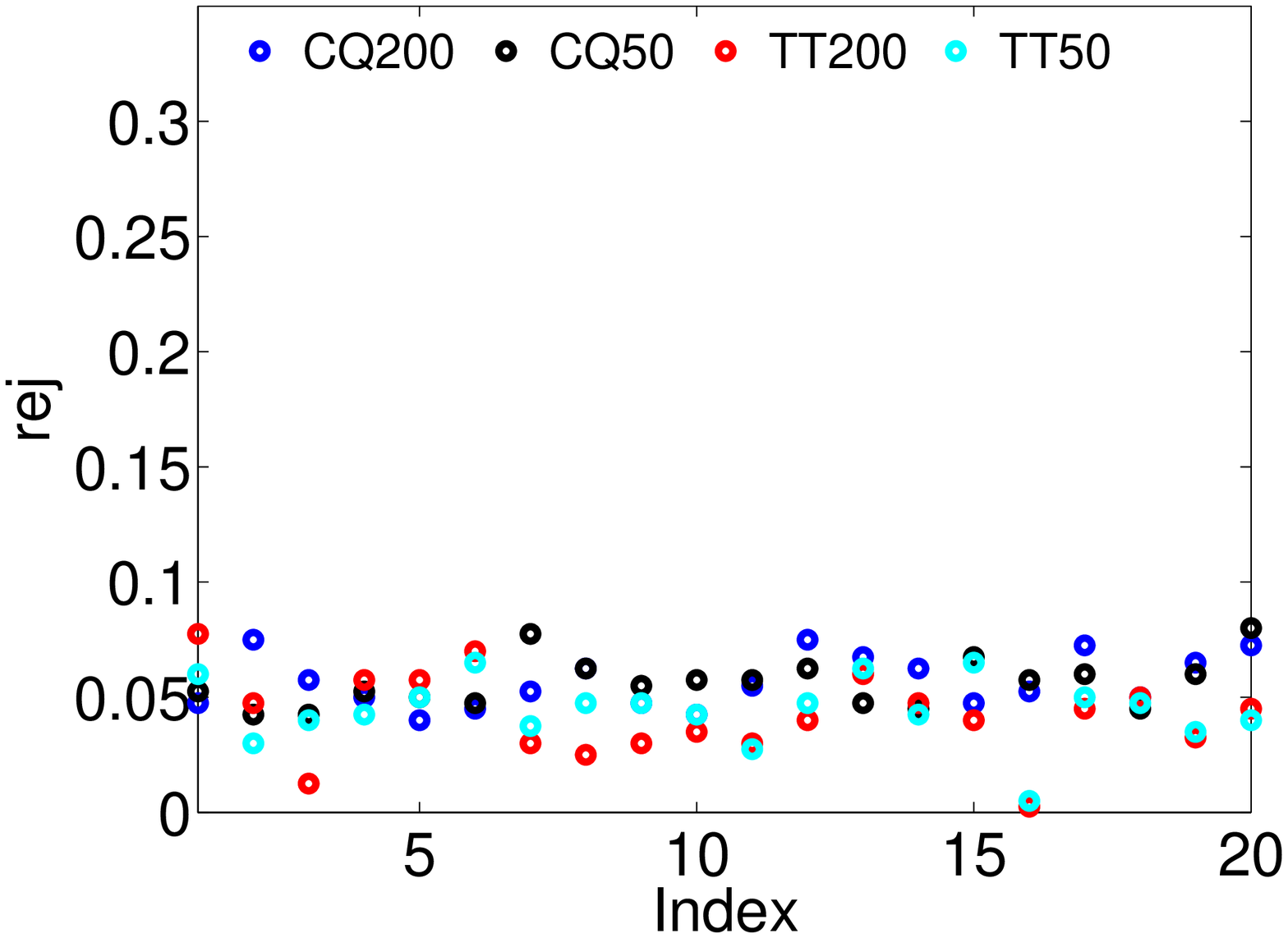}
\par\end{centering}
}\subfloat[Different]{\begin{centering}
\includegraphics[scale=0.21]{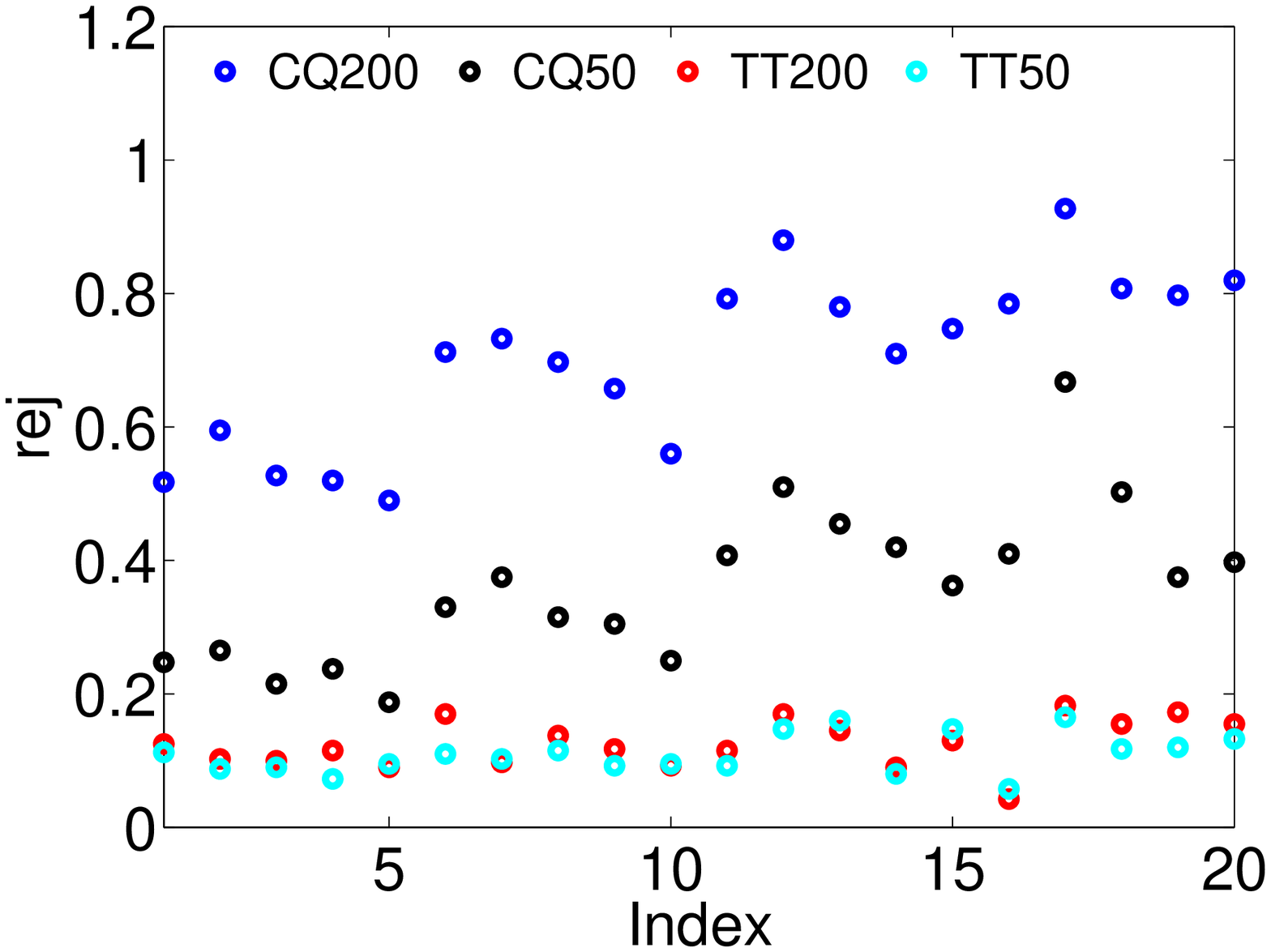}
\par\end{centering}
}\subfloat[Index]{\begin{centering}
\includegraphics[scale=0.21]{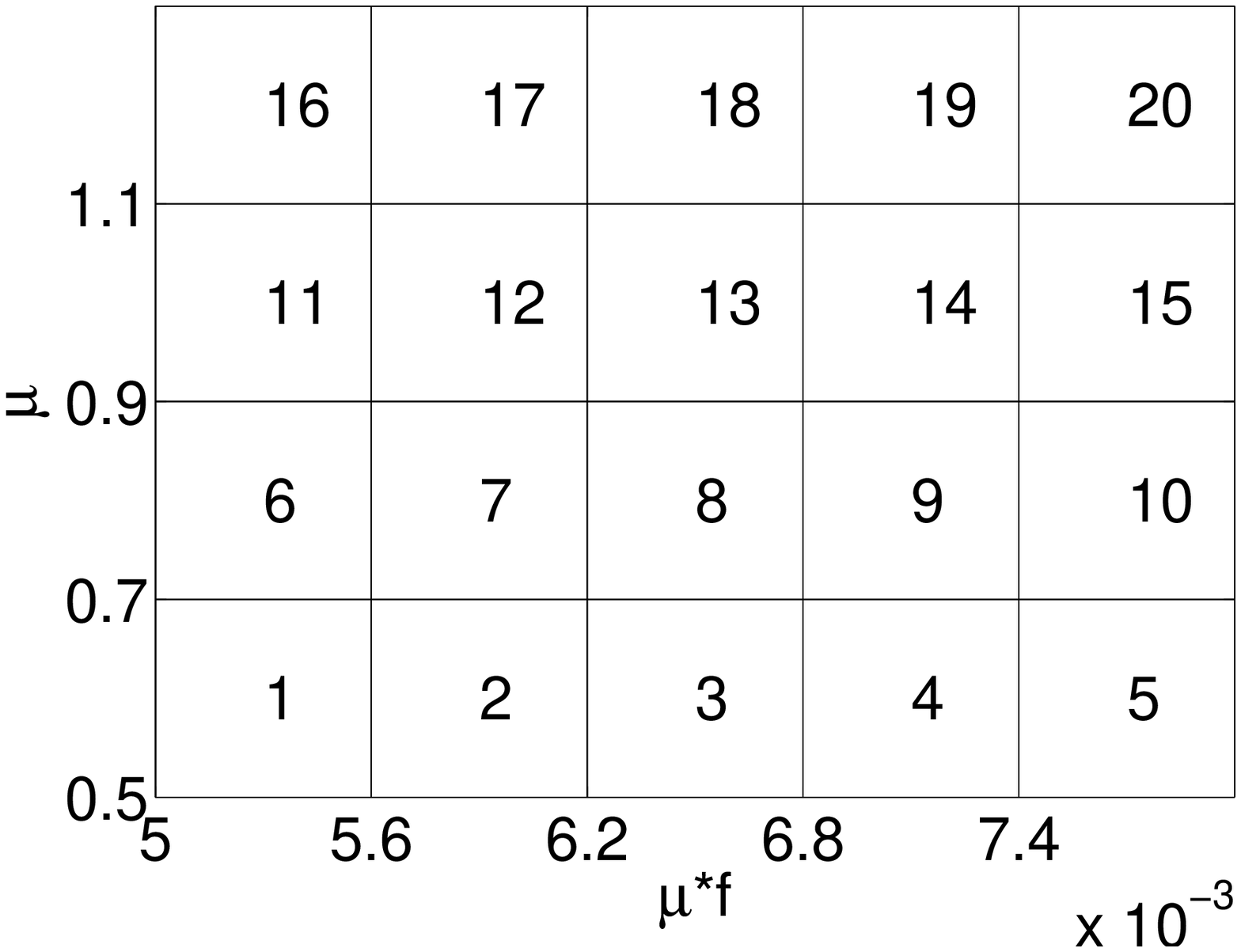}
\par\end{centering}
}
\par\end{centering}
\caption{{\bf Test results in different dynamical regimes of an I\&F neural network.}
The network consists of $160$ excitatory and $40$ inhibitory neurons
with random connections with probability $0.1$ driven by Poisson
inputs. The coupling strength between two connected neurons is selected
at random based on the uniform distribution of the interval $[0,s]$,
$s=0.01\unit{ms^{-1}}$ (the corresponding excitatory/inhibitory postsynaptic
potential is$\sim1\unit{mV}$). The Scanning parameters covers the
realistic firing rates ($\unit[3]{Hz}-\unit[50]{Hz}$) of real neurons.
The sampling time bin size is $\unit[0.5]{ms}$. $\Delta T$ is fixed
as $\unit[60]{ms}$. The recording time at every trial for the referential
and the test stimuli are $\unit[2]{s}$. We apply  two-sample test
methods for $400$ trials in four situations, CQ with $200$ neurons
(blue), CQ with $50$ neurons (black), TT with $200$ neurons (red)
and TT with $50$ neurons (cyan). For (a) and (b), the plot is the
rejection fraction. (a) The test stimuli are the same as the referential
one. (b) The test stimuli are $1.8\%$ larger in Poisson input magnitude
than the referential stimuli. The parameters of stimuli of each index
in (a) and (b) are from the left bottom corner of the box with the
corresponding index inside in (c). \label{fig:MDEPara1}}
\end{figure*}

Next, we examine whether this pretreatment is dependent of a particular
dynamical regime. Dynamical regimes are often realized by a particular
choice of network system parameters. We investigate this issue by
scanning the magnitude $f$ and the rate $\mu$ in the Poisson drive
of I\&F neural networks. The scanned range of these parameters produces
network dynamics with the range of firing rates  ($\unit[3]{Hz}-\unit[50]{Hz}$)
of real neurons. Note that there are typically three dynamical regimes
for the I\&F neurons with a fixed input magnitude $f$ \cite{zhou2014granger}:
(i) a highly fluctuating regime when the input rate $\mu$ is low;
(ii) an intermediate regime when $\mu$ is moderately high; (iii)
a low fluctuating or mean driven regime when $\mu$ is very high.
The underlying network consists of $160$ excitatory and $40$ inhibitory
neurons. It is a network of random connections with connection probability
$0.1$ and the coupling strength between two connected neurons is
selected at random based on the uniform distribution between $0$
and $1\unit{mV}$. Every trial requires $\unit[2]{s}$ recording time
for each stimulus. We also apply  two-sample test methods for $400$
trials in four situations, CQ with $200$ neurons (blue), CQ with
$50$ neurons (black), TT with $200$ neurons (red) and TT with $50$
neurons (cyan). Note that the $50$ neurons are also selected at random.
The parameters of stimuli of each index in Fig. \ref{fig:MDEPara1}a
and b are from the left bottom corner of the box with the corresponding
index inside in Fig. \ref{fig:MDEPara1}c. As is shown in Fig. \ref{fig:MDEPara1}a,
CQ and TT both work well for all scanned dynamical regimes, \emph{i.e.},
the rejection fractions of the null hypotheses $H_{0}$ for the case
of the same stimuli are around the selected significance level $0.05$.
When the test stimuli are $1.8\%$ larger in Poisson input magnitude
than those of the referential stimuli, Fig. \ref{fig:MDEPara1}b shows
that CQ has higher rejection fractions than those of TT, and CQ with
$200$ neurons has higher rejection fractions than those of CQ with
$50$ neurons. Conclusion is that, in all scanned dynamical regimes,
CQ method is better than TT method and CQ can use the advantage of
large neuron number. 

\subsection*{Swift discrimination tasks}

\begin{figure}
\begin{centering}
\subfloat[Same]{\begin{centering}
\includegraphics[scale=0.28]{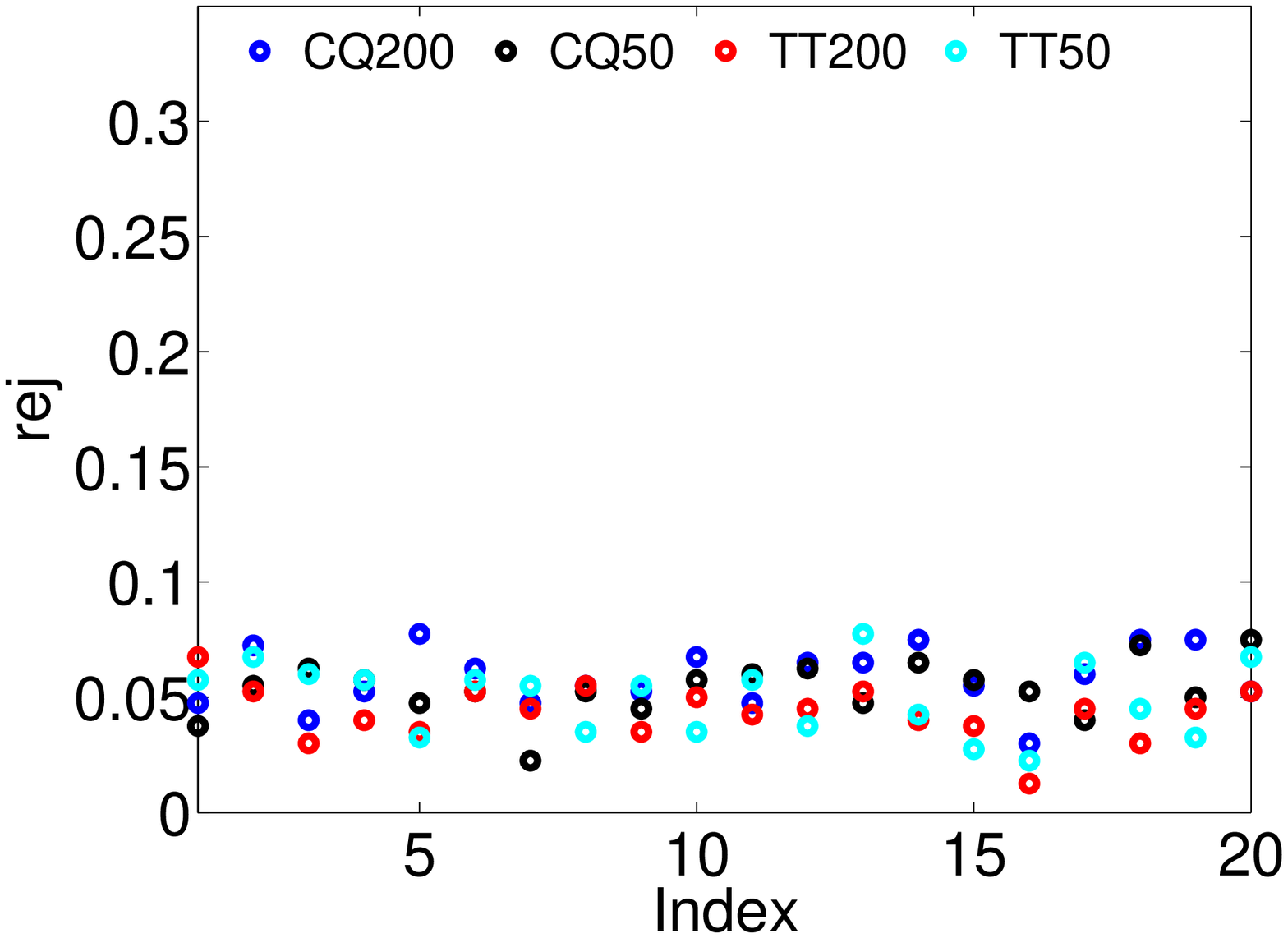}
\par\end{centering}
}\subfloat[Diff]{\begin{centering}
\includegraphics[scale=0.28]{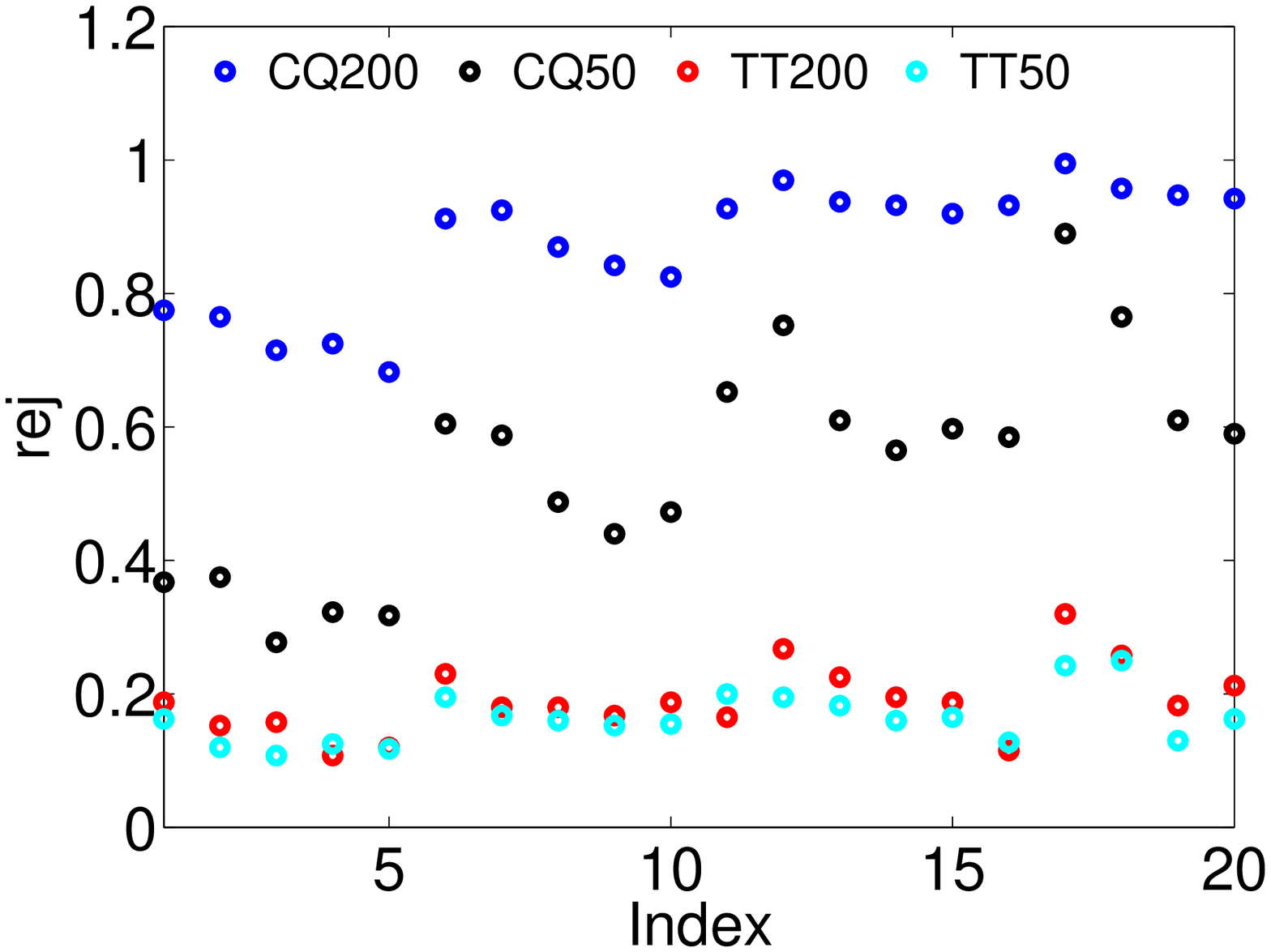}
\par\end{centering}
}
\par\end{centering}
\caption{{\bf Swift discrimination task. } The recording time at every trial
for $X_{1}$ and $X_{2}$ are $\unit[4]{s}$ and $\unit[360]{ms}$,
respectively. Other parameters and illustrations are the same as Fig.
\ref{fig:MDEPara1}. (a) The test stimuli are the same as the referential
one. (b) The test stimuli are $1.5\%$ larger in Poisson input magnitude
than the referential stimuli. \label{fig:MDEParaShort}}
\end{figure}

As is shown in above numerical examples, every trial requires $\unit[2]{s}$
recording time for each stimulus. However, a prey can react to a predator
in a shorter time, whereas it stays at a safe stimulus for a longer
time. We can mimic the prey to perform a discrimination task. Thus,
we can let the recording time of the referential stimulus $X_{1}$
longer, such as several seconds, and that of the test stimulus $X_{2}$
shorter, such as hundreds of milliseconds. There are usually two parts
in a test statistic, one is the summation of difference of mean values
{[}\emph{e.g. }Eq. (\ref{eq:Tnpart}){]}, the other is the variance
part {[}\emph{e.g. }Eqs. (\ref{eq:sig-1}), (\ref{eq:trs1new-1})
and (\ref{eq:trs12new-1}){]}, which is to normalize the statistic.
The variance estimator usually requires more data points, for example,
in our estimator for CQ statistic, the variance part requires at least
$6$ data points (See Methods.), whereas the mean difference part
requires at least $2$ data points. Since the recording time of $X_{1}$
is much longer than that of $X_{2}$, for CQ and TT methods, we can
estimate the variance part by $X_{1}$ solely under the null hypothesis
(See Methods.). We use a numerical example to show that this treatment
is able to deal with swift discrimination of hundreds of milliseconds.
Fig. \ref{fig:MDEParaShort} displays a swift discrimination version
of Fig. \ref{fig:MDEPara1}. The recording time at every trial for
$X_{1}$ and $X_{2}$ are $\unit[4]{s}$ and $\unit[360]{ms}$, respectively.
In the case of the same stimuli, as is shown in Fig. \ref{fig:MDEParaShort}a,
the rejection fractions of TT and CQ are around the selected significance
level $0.05$. This indicates TT and CQ are available to perform swift
discrimination tasks. In the case of different stimuli, in which the
test input is $1.5\%$ larger in Poisson input magnitude, the results
are similar as above results, \emph{i.e.}, as is shown in Fig. \ref{fig:MDEParaShort}b,
CQ is\textbf{ }better to discriminate different stimuli and CQ can
use advantage of large neuron number. 

\subsection*{Tuning curve}

Tuning curve is widely used to characterize the responses of one or
a group of neurons with respect to various inputs. The width of a
tuning curve is a popular way to potentially reflect the sensitivity
of considered neurons to the inputs. In general, a tuning curve of
firing rate is a classical and most used one. However, this idea is
based on one classical viewpoint, in which neurons are considered
to be independent. To understand how neurons code information, correlations
between neurons need to be probed. There is an idea \cite{samonds2003cooperation}
considering pairwise correlations, which we refer to as Dependency
in this paper (See Methods for details.). They consider the Kullback-Leibler
distance of joint distribution of two selected neurons to the distribution
of the forced-independent type, which is a type formed under the assumption
that the neurons are acting independently. 

To build a tuning curve by a two-sample test method, we establish
a quantity from the rejection fraction, 
\begin{equation}
N_{tst}=1-\rho,\label{eq:TSTtuning}
\end{equation}
where $\rho\in[0,1]$ is the rejection fraction as mentioned above,
it is larger for the input of larger difference. The quantity $N_{tst}$
is also in $[0,1]$, but it is smaller for the input of larger difference. 

\begin{figure}
\begin{centering}
\includegraphics[scale=0.45]{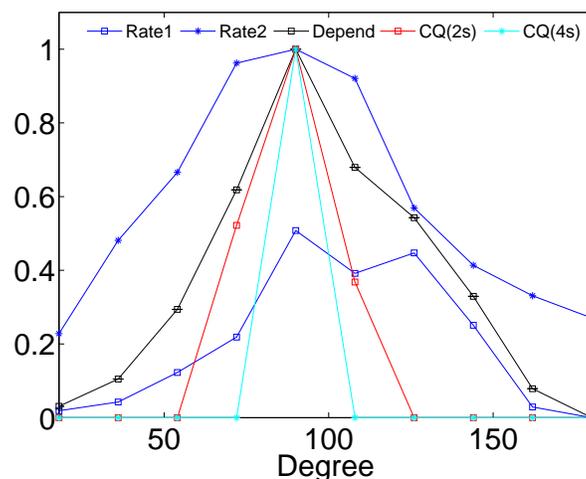}
\par\end{centering}
\caption{{\bf Normalized tuning curves in   orientation tasks.} We record
spike trains for drifting inputs of $10$ different orientations from
an I\&F neural network model of the primary visual cortex of the macaque
(See Methods for Details.). The sampling bin size is $\unit[0.5]{ms}$.
For the tuning curves of the firing rate and the Dependency, we selected
two neurons who have the same prefer orientations. The blue curves
are normalized tuning curves of the firing rate for the two selected
neurons, the black one is the normalized Dependency tuning curve of
the two selected neurons. The red and cyan ones are normalized tuning
curves of CQ method {[}Eq. (\ref{eq:TSTtuning}){]} using $\unit[2]{s}$
and $\unit[4]{s}$ data in each of $200$ trials, respectively, where
the referential orientation is the prefer orientation of the two selected
neurons in the method of Dependency. For tuning curves of CQ method,
we select about $340$ neurons to perform two-sample tests, where
the firing rates of the selected neurons are larger than $\unit[2.5]{Hz}$
at least under reference inputs or under test inputs. $\Delta T=\unit[60]{ms}$.
\label{fig:tuning}}
\end{figure}

To perform the comparison among different tuning curves, we record
spike trains for drifting inputs of $10$ different orientations from
an I\&F neural network model of the primary visual cortex of the macaque
(See Methods for Details.), as is shown in Fig. \ref{fig:tuning}.
For the tuning curves of the firing rate and the Dependency, we selected
two neurons who have the same prefer orientations. The blue curves
are normalized tuning curves of the firing rates for the two selected
neurons, the black one is the normalized Dependency tuning curve of
the two selected neurons. The red one is the normalized tuning curve
of CQ method {[}Eq. (\ref{eq:TSTtuning}){]} using $\unit[2]{s}$
recording data for both the referential and test stimuli in each of
$200$ trials, where the referential orientation is the prefer orientation
of the two selected neurons in the method of Dependency. The cyan
one is similar to the red one but using recording data of $\unit[4]{s}$.
For the tuning curves of CQ method, we select about $340$ neurons
to perform two-sample tests, where the firing rates of the selected
neurons are larger than $\unit[2.5]{Hz}$ under at least referenntial
inputs or test inputs. As is shown in Fig. \ref{fig:tuning}, the
curves of the CQ method are sharper than those of the other two methods,
which indicates the CQ method is potentially better than other methods
in discrimination tasks. The cyan curve is computed by CQ method with
$\unit[4]{s}$ data for each trial, which is sharper than the red
curve using $\unit[2]{s}$ data for each trial. This shows that the
CQ method is able to use the advantage of large neural number and
long recording time. It is no surprise that the CQ method is much
better than other methods since we are able to use more information
to estimate the turning curves.

\section*{Discussion }

To understand how neurons code stimulus in short time, a very first
step is to detect the difference of external stimuli in the activities
of neural networks. To achieve this goal, it requires proper high-dimensional
methods for neural data of ``large $p$, small $n$''. In another
aspect, there is great progress in statistics to deal with samples
of ``large $p$, small $n$''. However, because of neural dynamics,
the correlation of neural recordings over time forms a barrier between
the neural data and high-dimensional statistical methods. ACT method,
which transferring the correlation over time to the correlation over
dimensions, shows a possible way to overcome this barrier. Through
spike trains obtained from I\&F neural networks, we show that ACT
method enables the high-dimensional two-sample test methods to analyze
numerical neural data within very short time (hundreds of milliseconds).
Even there is only one trial data, with ACT method, we can also perform
a two-sample test on neural data with single-trial statistical power.

\begin{figure}
\begin{centering}
\includegraphics[scale=0.45]{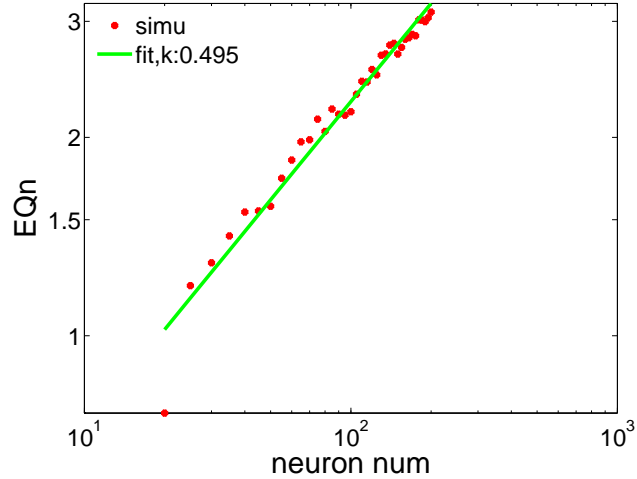}
\par\end{centering}
\caption{{\bf The relation between $\mathbf{ E(Q_n)}$ and  neuron number.}
This is one case in Fig. \ref{fig:MDEPara}, where $\mu=\unit[0.7]{\unit{ms^{-1}}}$
and $f=\unit[0.008]{\unit{ms^{-1}}}$. The red points are the $E(Q_{n})$
from numerical neural data with respect to the number of selected
neurons, and the green line is a linear fitting in the log-log plot
with slope $0.495$. This shows that $E(Q_{n})$ is a nearly linear
function of $\sqrt{p}$, where $p$ is neuron number.\label{fig:EQn-Neu}}
\end{figure}

Comparing to the TT method and other tuning curves considering only
few neurons, based on ACT method, the high-dimensional statistical
methods incorporating with neural correlations and neural number can
probe finer difference in the activities of neural networks. How well
the CQ method can use the advantage of large neuron number? As is
shown in Eq. (\ref{eq:EQn-1}) in Methods, in an example of data of
Gaussian distribution, since the covariance matrix is an identity,
the expectation $E(Q_{n})$ is a linear function of $\sqrt{p}$, where
$p$ is the dimension of the data. How is the situation in the neural
data? Fig. \ref{fig:EQn-Neu} displays one case of Fig. \ref{fig:MDEPara1},
where $\mu=\unit[0.7]{\unit{ms^{-1}}}$ and $f=\unit[0.008]{\unit{ms^{-1}}}$,
the expectation $E(Q_{n})$ is a nearly linear function of $\sqrt{p}$,
where $p$ is the selected neuron number. This linear relation is
a consequence of sparse covariance of neural data. 

The antidiagonal correlation is a limitation of extending our method
to data with slow decay dynamics. For the neural data, a $\Delta T$
of $\unit[60]{ms}$ also limits us from performing a two-sample test
on data of $\unit[100]{ms}$ recording time. 

\section*{Acknowledgments }

The authors thank Wei Dai for providing the code for the neural network
model of the macaque primary visual cortex. This work was supported
by NSFC-11671259, NSFC-11722107, and NSFC-91630208 (D.Z.); by NSFC-31571071
(D.C.); by Shanghai 14JC1403800, 15JC1400104, and SJTU-UM Collaborative
Research Program (D.C. and D.Z.); and by the NYU Abu Dhabi Institute
G1301 (Z.X., D.Z., and D.C.).

\section*{Methods}

\subsection*{High-dimensional two-sample test problem }

In general, we can describe the problem in the following way. $\mathbf{X}_{k1}$,$\mathbf{X}_{k2}$,$\cdots$,$\mathbf{X}_{kn_{k}}$$\in$$R^{p\times1}$
are a sample of data sampled independently from the $k$-th distribution,
where $k=1,2$, $p$ is the dimension and $n_{k}$ is the sample size
of the $k$-th sample. The mean value and covariance of the $k$-th
sample are denoted by $\mathbf{\mathbf{\mathbf{\mu}}}_{k}=(\mu_{k1},\mu_{k2},\cdots,\mu_{kp})^{T}\in R^{p\times1}$
and $\Sigma_{k}\in R^{p\times p}$, respectively. The null and alternative
hypotheses are as follows:
\begin{equation}
H_{0}:\mu_{1}=\mu_{2}\quad v.s.\quad H_{1}:\mu_{1}\neq\mu_{2},
\end{equation}
where the null hypothesis $H_{0}$ consists of $p$ marginal hypotheses,
\emph{i.e.}, $H_{0l}:\mu_{1l}=\mu_{2l}$, for $l=1,\cdots,p$. If
any component of $\mu_{1}$ is not equal to the corresponding one
of $\mu_{2}$, $H_{0}$ is rejected. The significance level for this
paper is selected as $0.05$.

\subsection*{CQ two-sample Test }

The CQ method \cite{chen2010two} considers the test statistic $Q_{n}$
as follows, under the null hypothesis,
\begin{equation}
Q_{n}\triangleq\frac{T_{n}}{\hat{\sigma}_{n}}\overset{d}{\rightarrow}N(0,1)\qquad p\rightarrow\infty,n\rightarrow\infty,\label{eq:Qn}
\end{equation}
where $\overset{d}{\rightarrow}$ indicates that the distribution
of $Q_{n}$ converges to the standard Gaussian distribution in distribution,
$T_{n}$ and $\hat{\sigma}_{n}$ are defined as below,
\begin{eqnarray}
T_{n} & = & \frac{\sum_{i\neq j}^{n_{1}}\mathbf{X}_{1i}^{T}\mathbf{X}_{1j}}{n_{1}(n_{1}-1)}+\frac{\sum_{i\neq j}^{n_{2}}\mathbf{X}_{2i}^{T}\mathbf{X}_{2j}}{n_{2}(n_{2}-1)}-\frac{2\sum_{i=1}^{n_{1}}\sum_{j=1}^{n_{2}}\mathbf{X}_{1i}^{T}\mathbf{X}_{2j}}{n_{1}n_{2}},\label{eq:Tnpart}
\end{eqnarray}
\begin{equation}
\hat{\sigma}_{n1}^{2}=\frac{2\widehat{tr(\Sigma_{1}^{2})}}{n_{1}(n_{1}-1)}+\frac{2\widehat{tr(\Sigma_{2}^{2})}}{n_{2}(n_{2}-1)}+\frac{4\widehat{tr(\Sigma_{1}\Sigma_{2})}}{n_{1}n_{2}},\label{eq:sig-1}
\end{equation}
in which $\hat{\sigma}_{n}$ is the standard deviation of $T_{n}$,
$\Sigma_{k}$ is the covariance of $\mathbf{X}_{2}$. We use the following
estimators for $\widehat{tr(\Sigma_{i}^{2})}$ and $\widehat{tr(\Sigma_{1}\Sigma_{2})}$,
which are slightly different with the estimators in Ref. \cite{chen2010two}.
We denote $A^{jk}$ is a copy of $X_{1}$ eliminating the $j$-th
and the $k$-th data point,
\begin{equation}
\widehat{tr(\Sigma_{1}^{2})}_{\mathrm{new}}=\{n_{1}(n_{1}-1)\}^{-1}tr\{\sum_{j\neq k}^{n_{1}}(\mathbf{X}_{1j}-\mathbf{c}_{11}^{jk})(\mathbf{X}_{1j}-\mathbf{c}_{12}^{jk})^{T}(\mathbf{X}_{1k}-\mathbf{c}_{13}^{jk})(\mathbf{X}_{1k}-\mathbf{c}_{14}^{jk})^{T}\},\label{eq:trs1new-1}
\end{equation}
where $\mathbf{c}_{1i}^{jk}$, $i=1,2,3,4$, is the sample mean of
the $i$-th quarter of $A^{jk}$. We denote $A^{l}$ is a copy of
$X_{1}$ eliminating the $l$-th sample and $B^{k}$ as a copy of
$X_{2}$ eliminating the $k$-th sample,
\begin{equation}
\widehat{tr(\Sigma_{1}\Sigma_{2})}_{\mathrm{new}}=\{n_{1}n_{2}\}^{-1}tr\{\sum_{l}^{n_{1}}\sum_{k}^{n_{2}}(\mathbf{X}_{1l}-\mathbf{d}_{11}^{l})(\mathbf{X}_{1l}-\mathbf{d}_{12}^{l})^{T}(\mathbf{X}_{2k}-\mathbf{d}_{21}^{k})(\mathbf{X}_{2k}-\mathbf{d}_{22}^{k})^{T}\},\label{eq:trs12new-1}
\end{equation}
where $\mathbf{d}_{1i}^{l}$, $i=1,2$ is the sample mean of the $i$-th
half of $A^{l}$, $\mathbf{d}_{2i}^{k}$, $i=1,2$ is the sample mean
of the $i$-th half of $B^{k}$. 

Elementary derivations show that
\begin{equation}
E(T_{n})=||\mu_{1}-\mu_{2}||^{2},\label{eq:ETnthe}
\end{equation}
where $E(x)$ is the expectation of a stochastic variable $x$, $E(X_{ki})=\mu_{k}$
for $k=1,2$ and $1\leq i\leq n_{k}$. 

The expectation of $Q_{n}$ is
\begin{equation}
E(Q_{n})=C\frac{n_{0}||\mu_{1}-\mu_{2}||^{2}}{\sqrt{tr(\tilde{\Sigma}^{2})}},\label{eq:EQn}
\end{equation}
where $n_{0}=n_{1}+n_{2}-2$, the composite covariance $\tilde{\Sigma}=(1-h)\Sigma_{1}+h\Sigma_{2}$,
$h=n_{1}/(n_{1}+n_{2})$ and 
\begin{equation}
C=\frac{h(1-h)}{\sqrt{2}}.\label{eq:EQC}
\end{equation}
To see the power of CQ method intuitively, we give a simple example.
Both samples have the same lengths, denoted as $n$, and the same
dimensions, denoted as $p$. The mean of the first sample is $\mu_{1}=0_{1\times p}$,
and that of the second sample is $\mu_{2}=1_{1\times p}$. Assuming
that the $k$-th sample ($k=1,2$) is sampled independently from the
normal distribution $N(\mu_{k},1)$, then, we have
\begin{equation}
E(Q_{n})\approx\frac{n||\mu_{1}-\mu_{2}||^{2}}{\sqrt{8tr(\Sigma^{2})}}=n\sqrt{p}/\sqrt{8}.\label{eq:EQn-1}
\end{equation}
 Since the expectation of $Q_{n}$ is crucial to the resolution of
the method, $p$ and $n$ are therefore important factors to the resolution.

In a swift two-sample test, the variance part is estimated by $X_{1}$
solely. Eq. (\ref{eq:sig-1}) would become 
\[
\hat{\sigma}_{n1}^{2}=\frac{2\widehat{tr(\Sigma_{1}^{2})}}{n_{1}(n_{1}-1)}+\frac{2\widehat{tr(\Sigma_{1}^{2})}}{n_{2}(n_{2}-1)}+\frac{4\widehat{tr(\Sigma_{1}^{2})}}{n_{1}n_{2}},
\]
 where $\widehat{tr(\Sigma_{1}^{2})}$ is estimated by Eq. (\ref{eq:trs1new-1}). 

\subsection*{Student t-test (TT) }

$\mathbf{x}=(x_{1},\cdots,x_{n_{x}})$ and $\mathbf{y}=(y_{1},\cdots,y_{n_{y}})$
are independently sampled from two one-dimensional distributions for
$n_{x}$ and $n_{y}$ times, respectively. The statistic quantity
of Student t-test (TT) test is defined as follows: 
\begin{equation}
Q_{n}^{TT}\triangleq\frac{\overline{x}-\overline{y}}{\sqrt{\frac{s_{x}^{2}}{n_{x}}+\frac{s_{y}^{2}}{n_{y}}}},\label{eq:QTT}
\end{equation}
where
\begin{equation}
\overline{x}=\frac{1}{n_{x}}\sum_{j=1}^{n_{x}}x_{j},
\end{equation}
\begin{equation}
\overline{y}=\frac{1}{n_{y}}\sum_{j=1}^{n_{y}}y_{j},
\end{equation}
\begin{equation}
s_{x}^{2}=\frac{1}{n_{x}-1}\sum_{j=1}^{n_{x}}(x_{j}-\overline{x})^{2},
\end{equation}
\begin{equation}
s_{y}^{2}=\frac{1}{n_{y}-1}\sum_{j=1}^{n_{y}}(y_{j}-\overline{y})^{2}.
\end{equation}
For data with dimension $p,$ it can be dealt with by TT one by one
dimension independently. In order to keep the total significance level
still being $0.05$, the significance level for each dimension is
selected by $0.05/p$. For each dimension, the acceptation of null
hypothesis is $1-0.05/p$. If we assume that data of different dimensions
are independent, for a large $p$, the acceptation of null hypothesis
of total test is 
\[
(1-\frac{0.05}{p})^{p}\approx1-0.05,
\]
which indicates the significance level for the whole test is still
around $0.05$.

In a swift two-sample test, the variance part is estimated by $x$
solely, \emph{i.e.}, $s_{y}^{2}$ is replaced by $s_{x}^{2}$ in the
$Q_{n}^{TT}$ in Eq. (\ref{eq:QTT}).

\subsection*{Integrate-and-fire model}

In this paper, we use the integrate-and-fire (I\&F for short) neuron
model for neural simulations. The dynamics of conductance-based I\&F
model \cite{gerstner2002spiking} is governed by 
\begin{equation}
\begin{cases}
\frac{dV_{i,Q}}{dt} & =-G^{L}(V_{i,Q}-\epsilon^{L})-G_{i,Q}^{E}(V_{i,Q}-\epsilon^{E})-G_{i,Q}^{I}(V_{i,Q}-\epsilon^{I}),\\
\frac{dG_{i,Q}^{E}}{dt} & =-\frac{G_{i,Q}^{E}}{\sigma^{E}}+S_{QE}\sum_{j_{E}\neq i}^{N_{E}}\sum_{k}m_{ij_{E}}\delta(t-T_{j_{E},k})+f_{E}\sum_{l}\delta(t-T_{i,l}^{F,E})\\
\frac{dG_{i,Q}^{I}}{dt} & =-\frac{G_{i,Q}}{\sigma^{I}}+S_{QI}\sum_{j\neq i}^{N_{I}}\sum_{k}m_{ij_{I}}\delta(t-T_{j_{I},k})+f_{I}\sum_{l}\delta(t-T_{i,l}^{F,I}),
\end{cases},\label{eq: IF 02}
\end{equation}
where the $i$-th neuron with type $Q=E$ or $I$, has both excitatory
conductance $G_{i,Q}^{E}$ and inhibitory conductance $G_{i,Q}^{I}$,
and the $\epsilon^{E}$ and $\epsilon^{I}$ are the excitatory and
inhibitory reversal potentials, respectively. $V_{i,Q}$ is the membrane
potential. $G^{L}$ is the leak conductance. The decay time scale
of excitatory and inhibitory synaptic conductance are $\sigma^{E}$
and $\sigma^{I}$, respectively. $S_{QE}$ and $S_{QI}$ are the strength
of input from excitatory and inhibitory neurons in the system, respectively.
We only use excitatory Poisson input to drive the system with the
rate $\mu$ and magnitude $f_{E}=f$, $f_{I}=0$. The network structure
is denoted by adjacency matrix $m_{ij}$.

The $i$-th neuron will evolve according to Eq. (\ref{eq: IF 02})
until $V_{i,Q}$ reaches the voltage threshold $V_{T}$ at which the
neuron will produce a spike. The spike time is denoted by $T_{i,k}^{F,Q}$,
where $i$ is the index of neuron, $k$ is the order of the spike
of neuron $i$. Then $V_{i,Q}$ is reset to be the resting potential
$V_{R}$ immediately. $V_{i,Q}$ will be kept at $V_{R}$ for a refractory
period $\tau_{ref}$ . When this is over, $V_{i,Q}$ starts to evolve
again.

The parameter set we used in the numerical experiment is: $\tau_{\mathrm{ref}}=\unit[2]{ms}$,
$\sigma^{E}=\unit[2]{ms}$, $\sigma^{I}=\unit[5]{ms}$, $G^{L}=\unit[0.05]{ms^{-1}}$.
The voltage uses a normalized unit: $V_{T}=1$, $V_{R}=0$, $\epsilon^{L}=0$,
$\epsilon^{E}=14/3$, $\epsilon^{I}=-2/3$. They correspond to typical
physiological values: $G^{L}=\unit[50\times10^{-6}]{\Omega^{-1}cm^{-2}}$,
$V_{T}=\unit[-55]{mV}$, $V_{R}=\epsilon^{L}=\unit[-70]{mV}$, $\epsilon^{I}=\unit[-80]{mV}$,
$\epsilon^{E}=\unit[0]{mV}$. We choose this set of parameters because
they are widely used.

\subsection*{Neural network model of the primary visual cortex of the macaque}

The simulation model of orientation discrimination task is an I\&F
neural network model of the primary visual cortex (V1) of the macaque,
details can be found at references \cite{tao2004egalitarian,tao2006orientation}.
In this paper, the network model consists of $4096$ neurons, $25\%$
inhibitory neurons. Each V1 cell sees a collection of Lateral geniculate
nucleus (LGN) neurons, where the number of LGN neurons is uniform
selected from $0$ to $30$. The on and off cells confer the orientation
and spatial phase preference of V1 cell. Orientation preference is
laid out in pinwheels. The V1 neurons also get inhibitory and excitatory
input from other V1 cells. The input is drifting sinusoid,
\[
I(\vec{x},t)=I_{0}[1+\epsilon\sin(\omega t-\vec{k}\cdot\vec{x}+\phi)],
\]
 where intensity $I_{0}=15.0$, contrast $\epsilon=100\%$, temporal
frequency $\omega=2\pi\text{\ensuremath{\times}}\unit[20]{Hz}$, and
spatial phase $\phi$ depends on the neural location. The spatial
frequency wave vector of the grating, $\vec{k}=k(\cos\theta,\sin\theta)$,
has spatial frequency $k=\unit[2\pi\text{\ensuremath{\times}}76.9]{\text{/cycle}}$
and orientation $\theta$. In the task, we pick an orientation input
as the referential stimuli, then, we can compute the rejection fraction
of a test orientation input by a two-sample test method. 

\subsection*{Tuning curve: Dependency}

Another tuning curve \cite{samonds2003cooperation} considers a pair
of neurons, which we refer to as Dependency. It considers the Kullback-Leibler
distance between joint distribution of two selected neurons and the
distribution of its forced-independent type, which is a type formed
under the assumption that the neurons are acting independently. By
denoting $P_{\sigma_{i}\sigma_{j}}$ is the probability of neuron
state $(\sigma_{i},\sigma_{j})$, the marginal distribution of first
neuron can be obtained by
\[
p_{1}^{1}=P_{10}+P_{11},
\]
\[
p_{0}^{1}=P_{01}+P_{00},
\]
 where $p_{\sigma}^{1}$ denotes the probability of the first neuron
with state $\sigma\in\{0,1\}$ of first neuron. Similarly, $p_{\sigma}^{2}$
can be evaluated. By assuming that neurons are independent, the joint
distribution of forced-independent type $Q_{\sigma_{i},\sigma_{j}}$
can be obtained,
\[
Q_{11}=p_{1}^{1}\times p_{1}^{2},
\]
\[
Q_{10}=p_{1}^{1}\times p_{0}^{2},
\]
 $Q_{01}$, $Q_{00}$ can be evaluated similarly. The Dependency is
defined as following,
\[
D=\sum_{\sigma_{i},\sigma_{j}\in\{0,1\}}P_{\sigma_{i},\sigma_{j}}\log\frac{P_{\sigma_{i},\sigma_{j}}}{Q_{\sigma_{i},\sigma_{j}}}.
\]

\section*{Appendix}

\subsection*{The relation between $\Sigma^{\prime}$ and $\Sigma$}

In ACT method, the time window $\Delta T$ is a free parameter. Nevertheless,
we show that the test result is not sensitive to the parameter $\Delta T$
when $\Delta T$ is larger than correlation length. To show this,
we prove an important relation Eq. (\ref{eq:newS}), which is a consequence
under the condition that the antidiagonal correlation is very weak.
For the sake of illustration, we consider a two-dimensional time series,
\[
\left(\begin{array}{c}
\mathbf{u}\\
\mathbf{v}
\end{array}\right)=\left(\begin{array}{ccccc}
u_{1} & u_{2} & \cdots & u_{i} & \cdots\\
v_{1} & v_{2} & \cdots & v_{i} & \cdots
\end{array}\right),
\]
where each dimension is an independent time series, the covariance
is denoted as 
\[
\Sigma=\left(\begin{array}{cc}
\sigma_{u}^{2} & \sigma_{uv}^{2}\\
\sigma_{uv}^{2} & \sigma_{v}^{2}
\end{array}\right).
\]
 If we select another $\Delta T^{\prime}=\alpha\Delta T$, for the
sake of simplicity, we assume that $\alpha$ is an integer, then,
we have another two-dimensional time series, 
\[
\left(\begin{array}{c}
\mathbf{U}\\
\mathbf{V}
\end{array}\right)=\left(\begin{array}{ccccc}
U_{1} & U_{2} & \cdots & U_{i} & \cdots\\
V_{1} & V_{2} & \cdots & V_{i} & \cdots
\end{array}\right),
\]
where 
\[
U_{i}=\frac{1}{\alpha}\sum_{j=1}^{\alpha}u_{(i-1)\alpha+j},
\]
\[
V_{i}=\frac{1}{\alpha}\sum_{j=1}^{\alpha}v_{(i-1)\alpha+j}.
\]
We denote the covariance of new time series as 
\[
\Sigma^{\prime}=\left(\begin{array}{cc}
\sigma_{U}^{2} & \sigma_{UV}^{2}\\
\sigma_{UV}^{2} & \sigma_{V}^{2}
\end{array}\right).
\]
The variance of $\mathbf{U}$ is 
\begin{equation}
\sigma_{U}^{2}=\mbox{var}(\frac{1}{\alpha}\sum_{j=1}^{\alpha}u_{j})=\frac{1}{\alpha}\sigma_{u}^{2}.\label{eq:SigU}
\end{equation}
The variance of $\mathbf{V}$ is 
\begin{equation}
\sigma_{V}^{2}=\mbox{var}(\frac{1}{\alpha}\sum_{j=1}^{\alpha}v_{j})=\frac{1}{\alpha}\sigma_{v}^{2}.\label{eq:SigV}
\end{equation}
The covariance of $\mathbf{U}$ and $\mathbf{V}$ is 
\begin{eqnarray*}
\sigma_{UV}^{2} & = & \mbox{\ensuremath{E}}[\frac{1}{\alpha}\sum_{i=1}^{\alpha}u_{i}-\mbox{\ensuremath{E}}(\frac{1}{\alpha}\sum_{j=1}^{\alpha}u_{j})][\frac{1}{\alpha}\sum_{i=1}^{\alpha}v_{i}-\mbox{E}(\frac{1}{\alpha}\sum_{j=1}^{\alpha}v_{j})]\\
 & = & \mbox{\ensuremath{E}}(\frac{1}{\alpha}\sum_{i=1}^{\alpha}u_{i}\cdot\frac{1}{\alpha}\sum_{j=1}^{\alpha}v_{j})-Eu_{1}Ev_{1}\\
 & = & \frac{1}{\alpha^{2}}\sum_{i,j=1}^{\alpha}E(u_{i}v_{j})-Eu_{1}Ev_{1},
\end{eqnarray*}
 where $\mbox{\ensuremath{E(x)}}$ denotes the expectation of a stochastic
variable $x$. We have a key assumption that the antidiagonal correlation
is very weak, \emph{i.e.}, 
\begin{equation}
E(u_{i}v_{j})\approx Eu_{i}Ev_{j},\quad i\neq j.\label{eq:indepAssumSig}
\end{equation}
Based on the assumption Eq. (\ref{eq:indepAssumSig}), we have 
\begin{eqnarray*}
\sigma_{UV}^{2} & = & \frac{1}{\alpha^{2}}\sum_{i=1}^{\alpha}E(u_{i}v_{i})+\frac{1}{\alpha^{2}}\sum_{i\neq j}^{\alpha}E(u_{i}v_{j})-Eu_{1}Ev_{1}\\
 & \approx & \frac{1}{\alpha}E(u_{1}v_{1})+\frac{\alpha(\alpha-1)}{\alpha^{2}}Eu_{1}Ev_{1}-Eu_{1}\mbox{E}v_{1}\\
 & = & \frac{1}{\alpha}(E(u_{1}v_{1})-Eu_{1}Ev_{1}),
\end{eqnarray*}
therefore, 
\begin{equation}
\sigma_{UV}^{2}\approx\frac{1}{\alpha}\sigma_{uv}^{2}.\label{eq:SigUV}
\end{equation}
From Eqs. (\ref{eq:SigU},\ref{eq:SigV},\ref{eq:SigUV}), we arrive
at 
\begin{equation}
\Sigma^{\prime}\approx\frac{1}{\alpha}\Sigma.\label{eq:Srelation}
\end{equation}

Since the covariance is a relation of every two dimensions, Eq. (\ref{eq:Srelation})
can be generalized to high-dimensional time series. The assumption
Eq. (\ref{eq:indepAssumSig}) is based on two facts, one is that the
correlation between neurons is small comparing to the variance \cite{Cohen2011Measuring},
the other is that we select a $\Delta T$ which is larger than the
correlation length of neuron data. Since no matter for $\Delta T$
or $\Delta T^{\prime}$, data processing is based on the original
sampled data, therefore, $\alpha$ does not have to be an integer.

\bibliographystyle{plos2015}
\bibliography{TST}

\end{document}